# Constructing and proving the ground state of a generalized Ising model by the cluster tree optimization algorithm

Wenxuan Huang[1], Daniil Kitchaev[1], Stephen Dacek[1], Ziqin Rong[1], Zhiwei Ding[1], Gerbrand Ceder[1,2,3]
1, Department of Material Science and Engineering, Massachusetts Institute of Technology, MA, USA
2, Department of Materials Science and Engineering, UC Berkeley, Berkeley, CA, USA
3, Materials Science Division, Lawrence Berkeley National Laboratory, Berkeley, CA, USA

## Abstract
Generalized Ising models, also known as cluster expansions, are an important tool in many areas of condensed-matter physics and materials science, as they are often used in the study of lattice thermodynamics, solid-solid phase transitions, magnetic and thermal properties of solids, and fluid mechanics. However, the problem of finding the global ground state of generalized Ising model has remained unresolved, with only a limited number of results for simple systems known.  We propose a method to efficiently find the periodic ground state of a generalized Ising model of arbitrary complexity by a new algorithm which we term cluster tree optimization. Importantly, we are able to show that even in the case of an aperiodic ground state, our algorithm produces a sequence of states with energy converging to the true ground state energy, with a provable bound on error. Compared to the current state-of-the-art polytope method, this algorithm eliminates the necessity of introducing an exponential number of variables to counter frustration, and thus significantly improves tractability. We believe that the cluster tree algorithm offers an intuitive and efficient approach to finding and proving ground states of generalized Ising Hamiltonians of arbitrary complexity, which will help validate assumptions regarding local vs. global optimality in lattice models, as well as offer insights into the low-energy behavior of highly frustrated systems.

## Introduction
The generalized Ising model[1], known to the materials science community as the cluster expansion[2-4], is the discrete representation of materials properties, e.g., formation energies, in terms of lattice sites and site interactions. It is a model widely applied to the study of configuration-property relationships[5-24], and has been an important tool in the study of, among others, magnetism [18], alloy thermodynamics [19], fluid dynamics[25], solid-solid phase transitions [20], and thermal conductivity [5]. One common application of cluster expansions is the determination of ground state structures and phase diagrams of crystalline solids based on a limited set of ab-initio calculations [8, 21-24] as the lowest energy states of a generalized Ising model determine the 0K phase diagram of the system. Therefore, a natural problem arises - given any set of effective cluster interactions (ECI's), or equivalently interaction parameters in the generalized Ising model, what is the exact ground state of the system?

The ground state problem is related to a well-studied problem in theoretical computer science - the Wang Tile problem[26-28]. The problem can be phrased as the following: given a set of squares with colored edges, is it possible for this set of tiles to tile a plane, with the constraint that neighboring edge colors must match. This problem has been shown to be undecidable[27, 28] due to the existence of sets of tiles for which only aperiodic tilings are admissible. Indeed, aperiodic solutions resulting from only 13 and 14 tiles have been constructed[29, 30]. It is worthwhile to note that edge-type Wang tiles could be converted to corner-type Wang tiles and vice versa[31, 32], implying that the tiling problem for the corner-type Wang tiles is also undecidable, although the smallest known set of aperiodic corner-type Wang tiles consists of 44 elements[32]. The undecidability of the Wang tile problem implies that the exact ground state problem which accounts for aperiodic states is similarly undecidable (see Supplementary Information), necessitating the use of approximate algorithms to solve the ground state problem in the most general case.

Currently, the most common approach to this problem is the Monte Carlo method, realized via the Metropolis algorithm[33], the Swendsen–Wang algorithm [34] and the Wolff algorithm[35]. However, the Monte Carlo approach does not provide a proof that the resulting low-energy configuration of the system is indeed the exact ground state given infinite degrees of freedom in spin variation. The traditional approach to find and prove the exact ground state is the polytope method introduced by Kaburagi and Kanamori [36, 37] combined with vertex enumeration[38]. The drawback of this method is that the constructed polytope has a exponential number of "unconstructible" vertices[39, 40] – combinations of correlation vectors, also known as lattice site "clusters", which do not correspond to any realizable lattice configuration - and despite recent advances in the field, there remains no general, tractable algorithm to obtain the true polytope.

In 2000, A. van de Walle demonstrated a way to generate valid inequalities for the polytope system to account for constructability and frustration[41]. However, this method is not guaranteed to produce all the necessary inequalities and thus is not guaranteed to converge the lower bound energy to the true ground state energy. More recently, Y.I Dublenych introduced a "basic rays" method to obtain the ground state of several small systems[42-44]. However, there is no known general algorithm based on this method, currently limiting its scope to simple model systems[42-44].

In this work, we present a general approach to the ground state problem, which we refer to as the "cluster tree optimization algorithm." We demonstrate that this algorithm is guaranteed to construct and prove, within an arbitrarily small numerical factor, the exact ground state for an arbitrary multicomponent set of ECIs on an arbitrary lattice system, assuming that a periodic ground state exists. We derive the algorithm by systematically constructing higher order polytopes without introducing exponentially many variables. Finally, we show that even in the case that the true ground state is aperiodic, our approach yields a series of converging spin configurations within an arbitrarily small margin of the true optimum. Compared with the state-of-the-art configurational polytope method[36, 37], our method moves from correlation space to appearance-frequency space. This conversion allows us to incrementally establish higher order configurational constraints, which is not possible in the traditional method. This conversion, together with a detailed implementation of the cluster tree algorithm, provides a systematic approach to deriving at the exact ground state of any cluster expansion.

Finally, we note that cluster tree optimization represents a useful procedure to approximate the generally undecidable Wang tile problem. Our method offers an effective procedure to determine tilability by converting the original problem into a series of efficient linear programming steps, providing a measure of tilability in the form of energy and a general direction to how the tiling could be constructed.

## Formalism

We begin by formally introducing the cluster tree optimization algorithm. We first show that for the purposes of cluster interactions, any lattice can be mapped to an orthorhombic multicomponent lattice without any symmetry in its interactions. We then prove that the total energy of this system can be written in terms of the energies of blocks of lattice sites. We proceed to define the basic polytope method for solving the ground state of such a system. Finally, we derive the "cluster tree optimization algorithm" and prove the correctness and generality of the method.

First, note that a binary generalized Ising Model on an arbitrary lattice with an arbitrary motif of n sites can always be represented by a generalized Ising model on an orthorhombic lattice with $2^n$ components without any symmetry. A proof and several examples of this transformation are given in the Supplementary Materials.

Second, we introduce the notion of a **"block"**. A block is a local configuration – for example, $(1,0,1,1)$ is a block in 1D binary system where a lattice site can be occupied by two species that we label "0" and "1". We define a **"minimal block"** as the smallest block that encapsulates all the interactions in the system – for example, $(0,0,0)$, $(0,0,1)$, and $(0,1,0)$ would all be minimal blocks for a 1D system with interaction up to the next nearest neighbor.

Third, we introduce the term **"energy of a block"** in order to use this change of basis for rewriting the Hamiltonian in terms of the **energies of blocks** and **appearance frequency of blocks**. As an example, consider the Ising Hamiltonian of a 1D lattice with nearest neighbor, next nearest neighbor, and triplet interactions. This Hamiltonian can be transformed into a sum over spin-configurations, multiplying of block energies by their appearance frequencies:

$$H = \mu \sum_{i \in \mathbb{Z}} \sigma_i + J_N \sum_{i \in \mathbb{Z}} \sigma_i \sigma_{i+1} + J_{NN} \sum_{i \in \mathbb{Z}} \sigma_i \sigma_{i+2} + J_{triplet} \sum_{i \in \mathbb{Z}} \sigma_i \sigma_{i+1} \sigma_{i+2}$$

$$= \sum_{i \in \mathbb{Z}} \left( \mu \sigma_i + J_N \sigma_i \sigma_{i+1} + J_{NN} \sigma_i \sigma_{i+2} + J_{triplet} \sigma_i \sigma_{i+1} \sigma_{i+2} \right)$$

$$= \sum_{i \in \mathbb{Z}} E\left(\sigma_i, \sigma_{i+1}, \sigma_{i+2}\right) = \sum_{(\sigma_1, \sigma_2, \sigma_3)} E\left(\sigma_1, \sigma_2, \sigma_3\right) \rho\left(\sigma_1, \sigma_2, \sigma_3\right)$$

where $\mu$ are the point energies, $J_N$ are the nearest neighbor interactions, $J_{NN}$ are the next nearest neighbor interactions, $J_{triplet}$ are the triplet interactions, $\sigma$ are spins, $\rho(\sigma_1, \sigma_2, \sigma_3)$ are the appearance frequencies for blocks $(\sigma_1, \sigma_2, \sigma_3)$, and $E(\sigma_1, \sigma_2, \sigma_3)$ are the block energies. To further illustrate the definition of appearance frequency, consider the periodic 1D configuration "---001001001001001---". In this configuration, the appearance frequencies would be $\rho[0] = \frac{2}{3}, \rho[1] = \frac{1}{3}, \rho[00] = \frac{1}{3}, \rho[01] = \frac{1}{3} \rho[10] = \frac{1}{3}, \rho[11] = 0$, and so on.

Similar arguments lead to results for 2D and 3D systems:

$$H = \sum_{(i,j) \in \mathbb{Z}^2} E \begin{pmatrix} \sigma_{i,j} & \cdots & \sigma_{i,j+M} \\ \vdots & \ddots & \vdots \\ \sigma_{i+N,j} & \cdots & \sigma_{i+N,j+M} \end{pmatrix} = \sum_{\{\sigma\}} E \begin{pmatrix} \sigma_{1,1} & \cdots & \sigma_{1,1+M} \\ \vdots & \ddots & \vdots \\ \sigma_{1+N,1} & \cdots & \sigma_{1+N,1+M} \end{pmatrix} \rho \begin{pmatrix} \sigma_{1,1} & \cdots & \sigma_{1,1+M} \\ \vdots & \ddots & \vdots \\ \sigma_{1+N,1} & \cdots & \sigma_{1+N,1+M} \end{pmatrix}$$

where the sum is over all possible configurations of $\{\sigma\}$. For the sake of brevity, we introduce a more compact notation:

$$\begin{pmatrix} \sigma_{i,j} & \cdots & \sigma_{i,j+M} \\ \vdots & \ddots & \vdots \\ \sigma_{i+N,j} & \cdots & \sigma_{i+N,j+M} \end{pmatrix} \equiv \sigma_{[i:i+N] \times [j:j+M]}$$

$$\begin{pmatrix} \sigma_{1,1} & \cdots & \sigma_{1,M} \\ \vdots & \ddots & \vdots \\ \sigma_{N,1} & \cdots & \sigma_{N,M} \end{pmatrix} \equiv \sigma_{[N] \times [M]}$$

which are adapted from mathematical convention that $[n] = \{1, 2, ..., n\}$ and $[i:i+M] = \{i, i+1, ..., i+M\}$. Thus, the Hamiltonian can be rewritten as:

$$H = \sum_{(i,j) \in \mathbb{Z}^2} E\left(\sigma_{[i:i+N] \times [j:j+M]}\right) = \sum_{\sigma_{[1+N] \times [1+M]}} E\left(\sigma_{[1+N] \times [1+M]}\right) \rho\left(\sigma_{[1+N] \times [1+M]}\right)$$

Based on these definitions, we can write down the **basic polytope method** for finding the ground state of an Ising Hamiltonian with a given set of interaction parameters. Consider a 2D system written in terms of block energies as described above, where all interactions fall within a range m by n. By defining the appearance frequency $\rho$ for each possible block, our objective is to:

$$\min_{\rho} \quad H[\{\sigma\}] = \sum_{\sigma_{[n] \times [m]}} \rho\left[\sigma_{[n] \times [m]}\right] E\left[\sigma_{[n] \times [m]}\right] \quad \text{(Eq. 1)}$$

However, constraints are needed on the $\rho$ variables in order for the solution to be physical. Thus, we introduce **compatibility equations** of order m by n as constraints for $\rho$. Formally, **compatibility equations** of order m by n are defined as the following equations - (Eq. 2), (Eq. 3), and (Eq. 4):

$$\rho\left[\sigma_{[n] \times [2:m]}\right] = \sum_{\sigma_{[n] \times [1]}} \rho\left[\begin{array}{cc} \sigma_{[n] \times [2:m]} & \sigma_{[n] \times [1]} \end{array}\right] = \sum_{\sigma_{[n] \times [1]}} \rho\left[\begin{array}{cc} \sigma_{[n] \times [1]} & \sigma_{[n] \times [2:m]} \end{array}\right] \quad \text{(Eq. 2)}$$

(Eq. 2) is a valid equality constraint on $\rho$ based on the simple observation that whenever $[\sigma]$ appears, its next neighbor must be either 0 or 1 (in the case of the binary system described earlier), corresponding to block $[\sigma 0]$ or block $[\sigma 1]$. Thus $\rho[\sigma] = \rho[\sigma 0] + \rho[\sigma 1]$, which is exactly the constraint given by (Eq. 2).

Furthermore, (Eq. 2) guarantees the **constructability** of $\rho$ in the x direction, where $\rho$ is deemed to be constructible if it corresponds to a physical lattice configuration, which will be proven later in this paper. A pictorial illustration of (Eq. 2) is shown in Figure 1a.

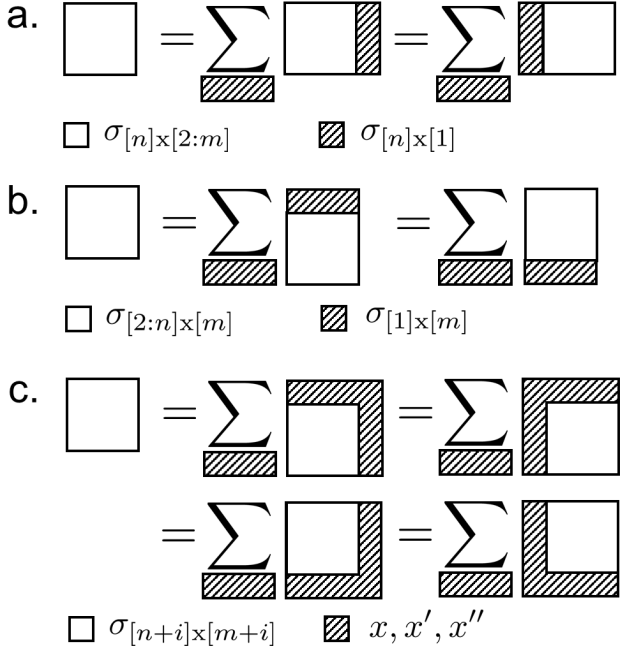

**Figure 1. a.** Pictorial illustration of the 1D block compatibility constraints defined in (Eq. 2), where the white block corresponds $\sigma_{[n]\times[2:m]}$ and the hatched block corresponds to $\sigma_{[n]\times[1]}$. **b.** Pictorial illustration of the 2D compatibility constraint added in (Eq. 3), where the white block corresponds $\sigma_{[2:n]\times[m]}$ and the hatched block corresponds to $\sigma_{[1]\times[m]}$. **c.** Pictorial illustration of the 2D perfect sum relationship, where the white block corresponds to a given [n+i] by [m+i] block, and the hatched block corresponds to all possible site configurations immediately adjacent to this block.

The next constraint is analogous, but given in the y-direction:

$$\rho\left[\sigma_{[2:n]\times[m]}\right] = \sum_{\sigma_{[1]\times[m]}} \rho\left[\begin{array}{c}\sigma_{[1]\times[m]}\\ \sigma_{[2:n]\times[m]}\end{array}\right] = \sum_{\sigma_{[1]\times[m]}} \rho\left[\begin{array}{c}\sigma_{[2:n]\times[m]}\\ \sigma_{[1]\times[m]}\end{array}\right] \quad \text{(Eq. 3)}$$

This constraint stems from similar reasoning as (Eq. 2) and guarantees constructability of $\rho$ in the y direction. A pictorial illustration is shown as in Figure 1b.

The final constraint we must add is that the set of all $\rho$ must correspond to a fully occupied lattice:

$$\sum_{\sigma_{[n]\times[m]}} \rho\left[\sigma_{[n]\times[m]}\right] = 1 \quad \text{(Eq. 4)}$$

The **basic polytope method** is formally defined as the linear programming minimization of (Eq. 1) subject to (Eq. 2), (Eq. 3), and (Eq. 4). Although we only show the formalism for 2D, the basic polytope method in 3D is exactly analogous.

Since every feasible solution must satisfy the compatibility equations, the linear system defined by these constraints provides a **lower bound** for the true ground state energy. Furthermore, one important result of this construction is that in any 1D problem, this lower bound is exact, meaning that any 1D problem can be fully solved by basic polytope method:

> **Proof:**
> Consider the interaction up to $n^{th}$ nearest neighbor, after transforming the Hamiltonian in terms of blocks:
> $$H = \sum_{\{\sigma\}} E(\sigma_1,\cdots,\sigma_{1+n})\rho(\sigma_1,\cdots,\sigma_{1+n})$$
> We could then construct a directed graph with all vertexes being of the form $(\sigma_1,\cdots,\sigma_n)$. Then, if $(\sigma_2,\cdots,\sigma_n) = (\sigma'_1,\cdots,\sigma'_{n-1})$, meaning the two blocks are off-by-one translations of each other, we associate an edge connecting $(\sigma_1,\cdots,\sigma_n)$ to $(\sigma'_1,\cdots,\sigma'_n)$ with a flow of size $\rho(\sigma_1,\cdots,\sigma_n,\sigma'_n)$.
> Note that in this system, each compatibility constraint is of the form:
> $$\rho(\sigma_1,\cdots,\sigma_n) = \sum_s \rho(\sigma_1,\cdots,\sigma_n,s) = \sum_s \rho(s,\sigma_1,\cdots,\sigma_n)$$
> meaning that in the directed graph, for each vertex, the sum of all the out-going flows from the vertex is equal to the sum of all in-coming flows into the vertex. By using the basic polytope method, one arrives at a flow solution $\rho$. Using analysis from linear programming and graph theory, specifically the network flow analysis[45], we know that this $\rho$ corresponds to a cycle in the directed graph and thus $\rho$ corresponds to a physical configuration. Thus, the ground state is given by such a configuration. ∎

However in two dimensions and higher, the polytope method thus defined fails in that it can give solutions that do not correspond to a real lattice configuration – we call these solutions **unconstructible**. The primary reason for this failure is that up to now, the constraints on the system guaranteed constructability in the x and y directions independently, not accounting for the fact that the x- and y- constructible solutions must also be compatible with each other. For example, the block configuration:

$$\rho\begin{pmatrix} 0 & 0 \\ 1 & 0 \end{pmatrix} = \rho\begin{pmatrix} 1 & 0 \\ 1 & 1 \end{pmatrix} = \rho\begin{pmatrix} 0 & 1 \\ 0 & 1 \end{pmatrix} = \rho\begin{pmatrix} 1 & 1 \\ 0 & 0 \end{pmatrix} = 0.25$$

satisfies the compatibility equations, but does not correspond to a real configuration on a lattice, making it an unconstructible solution. To be specific, $\begin{pmatrix} 0 & 0 \\ 1 & 0 \end{pmatrix}$ connects $\begin{pmatrix} 0 & 1 \\ 0 & 1 \end{pmatrix}$ to the left; $\begin{pmatrix} 0 & 1 \\ 0 & 1 \end{pmatrix}$ connects $\begin{pmatrix} 1 & 0 \\ 1 & 1 \end{pmatrix}$ to the left, but $\begin{pmatrix} 1 & 0 \\ 1 & 1 \end{pmatrix}$ does not connect to any other block cluster with non-zero appearing frequency to the left.

To account for constructability, we need a higher order polytope with additional constraints. Traditional approaches to this problem have relied on the enumeration of lattice configurations, which requires an exponential number of variables and makes the solution intractable. Instead, we introduce the cluster-tree optimization algorithm, which iteratively adds variables as necessary to counter frustration, reducing the prefactor in computational complexity to a more tractable level in practical cases and allows us to solve for the true, constructible ground state efficiently.

### Introducing the cluster tree optimization algorithm

The basic approach of our method is to converge an upper and lower bound on the ground state energy. First, we note that the energy of any spin configuration is trivially an upper bound on the ground state energy. Thus, we can obtain a tight upper bound by enumerating over potential periodicities and performing mixed integer programming minimization to obtain the lowest energy periodic solution within each choice of unit cell. However, without a tight lower bound on the energy, this calculation can never prove that any given solution is truly the ground state over all possible periodicities.

To find the lower bound, one could use the **basic polytope method**. If the lower bound does not converge to the energy obtained from the upper bound calculation, the lower bound can be refined by repeating the calculation with larger block sizes. In fact, it is possible to show that the lower bound obtained in this way converges to the true lower bound. This idea is in agreement with the traditional polytope method[36, 37], and suffers from the same problem - an exponential explosion of variables with respect to the block size. Thus, for a cluster expansion with an interaction range up to m by n, with the appearance frequency of a minimal block in the form $\rho\left[\sigma_{[n]\times[m]}\right]$ as in (Eq. 1), the objective of the **cluster tree optimization algorithm** is to obtain a refined lower bound as in the **basic polytope method** with a larger block size, $\rho\left[\sigma_{[n+K]\times[m+K]}\right]$, without generating all such variables. Instead, we generate only a few appearance frequency variables $\rho\left[\sigma_{[n+K]\times[m+K]}\right]$, such that a relationship, which we term the **perfect sum relationship**, similar to (Eq. 2), (Eq. 3), (Eq. 4) holds between the generated $\rho\left[\sigma_{[n+K]\times[m+K]}\right]$. This insight allows us to greatly reduce the variables present in the optimization problem, and thus improve the tractability of the lower bound optimization.

Formally, we define the **perfect sum relationship** holds for all block sizes below $[n+K] \times [m+K]$, if for any $\rho\left[\sigma_{[n+i] \times [m+i]}\right]$ where $i \in \{0, 1 \ldots K-1\}$

$$\rho\left[\sigma_{[n+i] \times [m+i]}\right] = \sum_{x,x',x''} \rho\begin{bmatrix} x'_{[1] \times [m+i]} & x'' \\ \sigma_{[n+i] \times [m+i]} & x_{[n+i] \times [1]} \end{bmatrix} = \sum_{x,x',x''} \rho\begin{bmatrix} \sigma_{[n+i] \times [m+i]} & x_{[n+i] \times [1]} \\ x'_{[1] \times [m+i]} & x'' \end{bmatrix}$$

$$= \sum_{x,x',x''} \rho\begin{bmatrix} x_{[n+i] \times [1]} & \sigma_{[n+i] \times [m+i]} \\ x'' & x'_{[1] \times [m+i]} \end{bmatrix} = \sum_{x,x',x''} \rho\begin{bmatrix} x'' & x'_{[1] \times [m+i]} \\ x_{[n+i] \times [1]} & \sigma_{[n+i] \times [m+i]} \end{bmatrix} \quad \text{(Eq. 5)}$$

where the sum is over all blocks that contains $\sigma_{[n+i] \times [m+i]}$ as its sub-block with a fixed cluster, shown pictorially in Figure 1c.

The $\rho\left[\sigma_{[n+K] \times [m+K]}\right]$ variables that are not generated are considered to be 0. In this way, without enumerating all $2^{|n+K|*|m+K|}$ variables, we could nonetheless obtain the refined lower bound at this block size. In the following section, we derive an algorithm that guarantees that the perfect sum relationship holds for all block sizes below $[n+K] \times [m+K]$, all the while generating the minimal possible number of configuration variables.

## Definition of the cluster tree optimization algorithm

The first step in the algorithm is to obtain an initial solution from the **basic polytope method**: minimize (Eq. 1) subject to the basic constraints given in (Eq. 2), (Eq. 3), and (Eq. 4). This solution is a first, loose lower bound of ground state energy.

To refine this lower bound (if possible), we need to introduce variables for the appearance frequency of larger blocks. We generate these variables using a **"spawning operation"**. A spawning operation on a variable, say $\rho[010]$, introduces a variable for the appearance frequency of a larger block, say $\rho[0100]$, such as to preserve the **perfect sum relationship** $\rho[010] = \rho[0100] + \rho[0101]$. To fully integrate the new $\rho[0100]$ variable, we add the necessary physical constraints $0 \leq \rho[0100] \leq \rho[100]$ and $0 \leq \rho[0101] \leq \rho[101]$ following the rules of the basic polytope method. Implicitly, this constraint $\rho[010] \leq \rho[100] + \rho[101]$ cuts out all unconstructible solutions where $\rho[010] \neq 0$, $\rho[100] = 0$ and $\rho[101] = 0$. Finally, after solving the new linear programming system, if we find that $\rho[0100] > 0$ and correspondingly, $\rho[100] > 0$ we introduce the **perfect sum relationship** constraint $\rho[0100] + \rho[1100] = \rho[100]$ into the linear programming system, which is a stronger condition than simply $\rho[0100] \leq \rho[100]$.

In 2D, the spawning operation follows the same concept but is more complex due to higher dimensionality and numerous possible shapes of the spawning block. However, as before, the spawning operation preserves the **perfect sum** relationship while introducing larger blocks into the linear programming system. The convergence and correctness of this approach will be proven in a later section. Here we present a brief summary of the 2D case and refer the reader interested in the exact derivation of the spawning operation to the supplementary material.

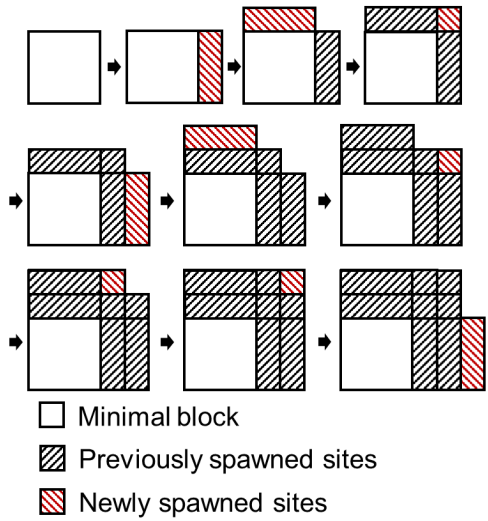

- ☐ Minimal block
- ▨ Previously spawned sites
- ▨ Newly spawned sites

**Figure 2.** An illustration of the Spawning procedure in 2D for the cluster tree optimization algorithm that generates blocks of increasing size while preserving the perfect-sum relationship and avoiding unnecessary variables. The hatched blocks indicate variables added to the original n by m block, where the red counter-hatched blocks specify the variables added in each specific spawning step.

Once again, consider a spin Hamiltonian in which all the interactions can be captured in a block of size m by n. As described earlier, we use a series of spawning operations to arrive at appearance frequencies of larger and larger blocks, giving us converging lower bounds on the total system energy. The general procedure for spawning is illustrated in Figure 2, where at each step, the red stars indicate the sites to be summed over. For example, the first iteration step illustrated in Figure 2 corresponds to the constraint:

$$\rho \begin{bmatrix} \sigma_{1,1} & \cdots & \sigma_{1,m} \\ \vdots & \ddots & \vdots \\ \sigma_{n,1} & \cdots & \sigma_{n,m} \end{bmatrix} = \sum_{*} \rho \begin{bmatrix} \sigma_{1,1} & \cdots & \sigma_{1,m} & * \\ \vdots & \ddots & \vdots & \vdots \\ \sigma_{n,1} & \cdots & \sigma_{n,m} & * \end{bmatrix}$$

Note that Figure 2 only demonstrates one direction of spawning, while in reality there are 3 other spawning directions as illustrated in Figure 1c. The other spawning directions can be derived by exact analogy to the procedure described above.

An essential detail to any spawning operation is that before spawning a variable of the form $\rho\left[\sigma_{[n+i]\times[m+i]}\right]$ with $i>0$, one needs to ensure that the perfect sum relationship holds for all block sizes below $[n+i]\times[m+i]$. Thus, for all $\rho\left[\sigma_{[n+i]\times[m+i]}\right]$, we need to ensure that $\rho\left[\sigma_{[n+i]\times[m+i]-\{(1,1)\}}\right]$, $\rho\left[\sigma_{[n+i]\times[m+i]-\{(n+i,1)\}}\right]$, $\rho\left[\sigma_{[n+i]\times[m+i]-\{(1,m+i)\}}\right]$, $\rho\left[\sigma_{[n+i]\times[m+i]-\{(n+i,m+i)\}}\right]$ have been generated in the calculation, so that we can impose the constraints:

$$\rho\left[\sigma_{[n+i]\times[m+i]}\right] \leq \rho\left[\sigma_{[n+i]\times[m+i]-\{(1,1)\}}\right]$$

$$\rho\left[\sigma_{[n+i]\times[m+i]}\right] \leq \rho\left[\sigma_{[n+i]\times[m+i]-\{(1,m+i)\}}\right]$$

$$\rho\left[\sigma_{[n+i]\times[m+i]}\right] \leq \rho\left[\sigma_{[n+i]\times[m+i]-\{(n+i,1)\}}\right]$$

$$\rho\left[\sigma_{[n+i]\times[m+i]}\right] \leq \rho\left[\sigma_{[n+i]\times[m+i]-\{(n+i,m+i)\}}\right] \quad \text{(Eq. 6)}$$

We refer this process as **adding maximal constraints**. Having introduced the maximal constraints and solved the linear optimization again, if $\rho\left[\sigma_{[n+i]\times[m+i]}\right]>0$, we can finally establish the constructability constraints:

$$\sum_{\sigma_{(1,1)}} \rho\left[\sigma_{[n+i]\times[m+i]}\right] = \rho\left[\sigma_{[n+i]\times[m+i]-\{(1,1)\}}\right]$$

$$\sum_{\sigma_{(n+i,1)}} \rho\left[\sigma_{[n+i]\times[m+i]}\right] = \rho\left[\sigma_{[n+i]\times[m+i]-\{(n+i,1)\}}\right]$$

$$\sum_{\sigma_{(1,m+i)}} \rho\left[\sigma_{[n+i]\times[m+i]}\right] = \rho\left[\sigma_{[n+i]\times[m+i]-\{(1,m+i)\}}\right]$$

$$\sum_{\sigma_{(n+i,m+i)}} \rho\left[\sigma_{[n+i]\times[m+i]}\right] = \rho\left[\sigma_{[n+i]\times[m+i]-\{(n+i,m+i)\}}\right] \quad \text{(Eq. 7)}$$

However, following the spawning procedure illustrated in Figure 2, it is possible that some of $\rho\left[\sigma_{[n+i]\times[m+i]-\{(1,1)\}}\right]$, $\rho\left[\sigma_{[n+i]\times[m+i]-\{(n+i,1)\}}\right]$, $\rho\left[\sigma_{[n+i]\times[m+i]-\{(1,m+i)\}}\right]$, $\rho\left[\sigma_{[n+i]\times[m+i]-\{(n+i,m+i)\}}\right]$ variables are not generated when $\rho\left[\sigma_{[n+i]\times[m+i]}\right]$ needs to be spawned. Without loss of generality, suppose the missing block is $\rho\left[\sigma_{[n+i]\times[m+i]-\{(1,m+i)\}}\right]$. In this case, we need to trace back in Figure 2 to find the closest block $\sigma'$ that has already been generated, and impose the constraint:

$$\rho\left[\sigma_{[n+i]\times[m+i]}\right] \leq \rho\left[\sigma'\right]$$

We define this process as **back tracing**. So long as $\rho\left[\sigma_{[n+i]\times[m+i]}\right] > 0$ in subsequent computations, $\rho[\sigma'] > 0$ holds and $\rho[\sigma']$ can be back traced to eventually yield all the missing blocks.

To summarize, if the algorithm is about to spawn $\rho\left[\sigma_{[n+i]\times[m+i]}\right]$, one needs to first either immediately add **maximal constraints** or **back trace** to ensure that $\rho\left[\sigma_{[n+i]\times[m+i]}\right]$ preserves the perfect sum relationship.

With **basic polytope method, spawning,** and adding **maximal constraints** defined, the pseudo code of the cluster tree optimization algorithm is as follows:
1. Use the basic polytope method to initiate a linear programming system to obtain the appearing frequency of minimal blocks
2. Collect the set of blocks with the smallest size and a positive appearing frequency, denote the set by $S$
3. If all elements of $S$ is in the form $\sigma_{[n+i]\times[m+i]}$ for some $i > 0$, then
    a. If for all $\sigma \in S$, the maximal constraints for $\rho[\sigma]$ have been added, spawning $\rho[\sigma]$ for all $\sigma \in S$ to generate a new set of larger blocks.
    b. Otherwise, try to add maximal constraints for $\rho[\sigma]$ for all $\sigma \in S$ either directly, or by back tracing.
4. Solve the linear programming system to obtain the refined lower bound and repeat from step 2.

The optimization loop terminates when either the computed lower bound matches the previously calculated upper bound, or when the spawning size *i* reaches some maximum defined threshold *N*.

When the cluster tree optimization algorithm terminates, if the lower bound and upper bound match, we can guarantee that the ground state solution has been found. Otherwise, based on the fact that the **perfect sum relationship** holds for all blocks with size below $[n+N]\times[m+N]$, we arrive at a converging lower bound as N increases. In practical cases, we find that this convergence tends to be finite, meaning that the lower bound matches the upper bound after some finite number of iterations, as spawning directly corresponds to establishing larger and larger clusters in the traditional polytope method. However, this general finite convergence property cannot be proved.

In this method, we have introduced variables corresponding to interactions of a much higher order than those present in the original problem. Nonetheless, the performance of this approach is vastly superior to direct enumeration as required by traditional methods. The traditional polytope method in general requires $2^{n^2}$ variables in the binary case, or $k^{n^2}$ variables in k-nary case, to account for clusters of size n by n, while for this method such exponentiation is not necessary. For example, we find that to solve a system with a maximum cluster size of 10 by 10, our method requires approximately 50,000 variables, compared to the completely intractable $2^{100}$ variables needed for direct enumeration.

As a final observation, there is one alternative termination condition for the optimization. If the algorithm reaches step 3.a with the $i$ defined in that step, meaning that maximal constraints has been added for all $\sigma_{[n+i] \times [m+i]}$ such that $\rho\left[\sigma_{[n+i] \times [m+i]}\right] > 0$, and if all such $\sigma_{[n+i] \times [m+i]}$ admit the same periodicity, then we immediately know that the current lower bound is the true ground state energy and $\rho$ is constructible. The proof of this termination condition is given in the supplementary information.

## Results

Having defined the cluster-tree optimization algorithm, we illustrate that our solver can reproduce and prove the correctness of ground states known in the literature [46]. In the following examples we look at a triangular lattice with interactions up to the third nearest neighbor. The first step is to define bijection between a triangular and square lattice by setting $(1,0) \to [1,0]$ and $\left(\frac{1}{2}, \frac{\sqrt{3}}{2}\right) \to [0,1]$, as shown in Figure 3.

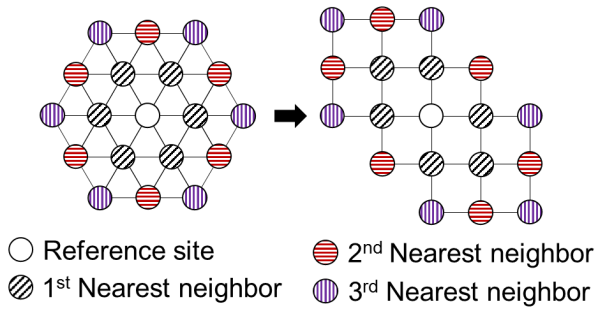

○ Reference site  ⊜ 2nd Nearest neighbor
⊘ 1st Nearest neighbor  ⦀ 3rd Nearest neighbor

**Figure 3. Mapping** the **interaction**s on **a triangular lattice** to an equivalent set on **a square lattice**, with loss of symmetry **in the interactions.**

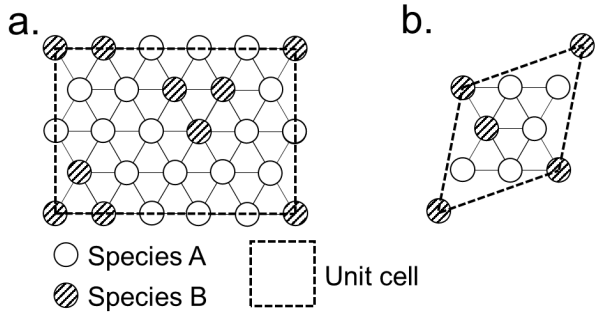

**Figure 4 a.** The known ground state structure of a pair-interaction Hamiltonian with $V_0 = -4, V_1 = 1, V_2 = 1, V_3 = 1$, where $V_0, V_1, V_2, V_3$ corresponds to the point term, nearest neighbor, next nearest neighbor and 3rd nearest neighbor interaction terms. **b.** Known ground state of the frustrated Hamiltonian with $V_1 = 2, V_2 = 1, V_3 = 1, V_0 = -6$, where $V_0, V_1, V_2, V_3$ are defined in the same way.

**Example 1**: It is known that the structure in Figure 4a corresponds to the ground state of with the interaction parameters $V_0 = -4 \quad V_1 = 1 \quad V_2 = 1 \quad V_3 = 1$, where $V_0, V_1, V_2, V_3$ correspond to the point term, nearest neighbor, next nearest neighbor and 3rd nearest neighbor interaction terms on a triangular lattice. Using only the basic polytope method and periodicity enumeration, we can already prove the ground state on an equivalent square lattice. Clearly, in the most basic cases, the polytope method can immediately yield a converged lower bound on the energy. The reason for this success is that this particular Hamiltonian is not frustrated. In the next example, we consider a frustrated system to see how the cluster-tree optimization algorithm efficiently counters frustration, giving a superior result to the basic polytope method.

**Example 2:** It is known that the ground state corresponding to interaction parameters $V_1 = 2, V_2 = 1, V_3 = 1, V_0 = -6$, where V are defined as before, is the given in structure in Figure 4b[37]. From periodicity enumeration, the ground state energy is suggested to be -1.143, yielding a structure symmetrically equivalent to the true ground state shown in Figure 4b. However, the basic polytope method produces a lower bound of -1.153, which does not match the energy obtained from site enumeration. The cluster tree algorithm in the other hand yields a lower bound energy of -1.143 after 4 iterations, consistent with that provided by this ground state structure.

**First iteration:** From the basic polytope method equation, we calculate blocks with non-zero appearing frequency to be:

$$\rho\begin{bmatrix}000\\000\\011\end{bmatrix}=\rho\begin{bmatrix}000\\001\\001\end{bmatrix}=\rho\begin{bmatrix}000\\011\\100\end{bmatrix}=\rho\begin{bmatrix}100\\000\\001\end{bmatrix}=\rho\begin{bmatrix}100\\100\\000\end{bmatrix}=\rho\begin{bmatrix}100\\001\\010\end{bmatrix}=\rho\begin{bmatrix}010\\100\\001\end{bmatrix}$$

$$=\rho\begin{bmatrix}010\\010\\100\end{bmatrix}=\rho\begin{bmatrix}110\\000\\000\end{bmatrix}=\rho\begin{bmatrix}001\\010\\010\end{bmatrix}=\rho\begin{bmatrix}001\\110\\000\end{bmatrix}=\rho\begin{bmatrix}001\\001\\110\end{bmatrix}=\rho\begin{bmatrix}011\\100\\100\end{bmatrix}=\frac{1}{13}$$

(Eq. 8)

**Second iteration:** All the non-zero blocks are spawned in the horizontal direction after adding maximal constraints to the system. For example:

$$\rho\begin{bmatrix}000\\000\\011\end{bmatrix}=\rho\begin{bmatrix}0000\\0000\\0110\end{bmatrix}+\rho\begin{bmatrix}0001\\0000\\0110\end{bmatrix}+\rho\begin{bmatrix}0000\\0001\\0110\end{bmatrix}+\rho\begin{bmatrix}0001\\0001\\0110\end{bmatrix}+\rho\begin{bmatrix}0000\\0000\\0111\end{bmatrix}+\rho\begin{bmatrix}0001\\0000\\0111\end{bmatrix}+\rho\begin{bmatrix}0000\\0001\\0111\end{bmatrix}+\rho\begin{bmatrix}0001\\0001\\0111\end{bmatrix}$$

$$\rho\begin{bmatrix}000\\000\\110\end{bmatrix}\geq\rho\begin{bmatrix}0000\\0000\\0110\end{bmatrix}\quad\rho\begin{bmatrix}001\\000\\110\end{bmatrix}\geq\rho\begin{bmatrix}0001\\0000\\0110\end{bmatrix}\quad\rho\begin{bmatrix}000\\001\\110\end{bmatrix}\geq\rho\begin{bmatrix}0000\\0001\\0110\end{bmatrix}....$$

With these new constraints, Solving linear programming reproduces equation (Eq. 8) and generates (Eq. 9).

$$\rho\begin{bmatrix}0001\\0001\\0110\end{bmatrix}=\rho\begin{bmatrix}0001\\0010\\0010\end{bmatrix}=\rho\begin{bmatrix}0001\\0110\\1000\end{bmatrix}=\rho\begin{bmatrix}1000\\0000\\0011\end{bmatrix}=\rho\begin{bmatrix}1000\\1001\\0001\end{bmatrix}=\rho\begin{bmatrix}1000\\0011\\0100\end{bmatrix}=\rho\begin{bmatrix}0100\\1001\\0010\end{bmatrix}$$

$$=\rho\begin{bmatrix}0100\\0100\\1000\end{bmatrix}=\rho\begin{bmatrix}1100\\0000\\0001\end{bmatrix}=\rho\begin{bmatrix}0011\\0100\\0100\end{bmatrix}=\rho\begin{bmatrix}0010\\1100\\0001\end{bmatrix}=\rho\begin{bmatrix}0010\\0010\\1100\end{bmatrix}=\rho\begin{bmatrix}0110\\1000\\1000\end{bmatrix}=\frac{1}{13}$$

(Eq. 9)

**Third iteration:** We then spawn all those nonzero blocks that have not been previously spawned, for example:

$$\rho\begin{bmatrix}0011\\0100\\0100\end{bmatrix}=\sum_{i,j,k}\rho\begin{bmatrix}ijkx\\0011\\0100\\0100\end{bmatrix}$$

$$\rho\begin{bmatrix}101x\\0011\\0100\\0100\end{bmatrix}\leq\rho\begin{bmatrix}101\\001\\010\end{bmatrix}\quad\ldots$$

where $x$ is could be simply thought as empty space to make its representation clearer. In this step, linear programming results in (Eq. 8), (Eq. 9) and (Eq. 10) where the x in equation 8 refers to an empty space. There are all together 52 such terms, where we only give a representative sample:

$$\rho\begin{bmatrix}110x\\0001\\0001\\0110\end{bmatrix}=\rho\begin{bmatrix}x000\\0001\\0001\\0110\end{bmatrix}=\rho\begin{bmatrix}0001\\0001\\0110\\x000\end{bmatrix}=\rho\begin{bmatrix}0001\\0001\\0110\\100x\end{bmatrix}$$

$$=\rho\begin{bmatrix}100x\\1000\\0000\\0011\end{bmatrix}=\rho\begin{bmatrix}x110\\1000\\0000\\0011\end{bmatrix}=\rho\begin{bmatrix}1000\\0000\\0011\\x100\end{bmatrix}=\rho\begin{bmatrix}1000\\0000\\0011\\001x\end{bmatrix}$$

$$=\rho\begin{bmatrix}000x\\0110\\1000\\1000\end{bmatrix}=\rho\begin{bmatrix}x001\\0110\\1000\\1000\end{bmatrix}=\rho\begin{bmatrix}0110\\1000\\1000\\x011\end{bmatrix}=\rho\begin{bmatrix}0110\\1000\\1000\\000x\end{bmatrix}=\cdots=\frac{1}{13}$$

(Eq. 10)

**Forth iteration:** This step is crucial in countering the frustration effect. Again, every non-zero block is spawned. The most important of these for countering frustration in the system is:

$$\rho\begin{bmatrix}0110\\1000\\1000\\000x\end{bmatrix}=\rho\begin{bmatrix}0110\\1000\\1000\\0000\end{bmatrix}+\rho\begin{bmatrix}0110\\1000\\1000\\0001\end{bmatrix}$$

$$\rho\begin{bmatrix}0110\\1000\\1000\\0000\end{bmatrix}\le\rho\begin{bmatrix}000\\000\\000\end{bmatrix}$$

$$\rho\begin{bmatrix}0110\\1000\\1000\\0001\end{bmatrix}\le\rho\begin{bmatrix}000\\000\\001\end{bmatrix}$$

Note that neither $\rho\begin{bmatrix}000\\000\\000\end{bmatrix}$ nor $\rho\begin{bmatrix}000\\000\\001\end{bmatrix}$ is larger than 0 in the previous solution in equation 6, but the spawned term $\rho\begin{bmatrix}0110\\1000\\1000\\000x\end{bmatrix}>0$ from equation 8, meaning that that either one of $\rho\begin{bmatrix}000\\000\\000\end{bmatrix}$ or $\rho\begin{bmatrix}000\\000\\001\end{bmatrix}$ must be larger than 0.

Thus the next linear programming calculation forces $\rho\begin{bmatrix}0110\\1000\\1000\\000x\end{bmatrix}=0$ or $\rho\begin{bmatrix}000\\000\\000\end{bmatrix}>0$ or $\rho\begin{bmatrix}000\\000\\001\end{bmatrix}>0$.

Solving the linear system again, a brand new solution is obtained and the frustration effect has been countered:

$$\rho\begin{bmatrix}000\\010\\001\end{bmatrix}=\rho\begin{bmatrix}000\\101\\000\end{bmatrix}=\rho\begin{bmatrix}100\\010\\000\end{bmatrix}=\rho\begin{bmatrix}010\\000\\101\end{bmatrix}=\rho\begin{bmatrix}010\\001\\100\end{bmatrix}=\rho\begin{bmatrix}001\\100\\010\end{bmatrix}=\rho\begin{bmatrix}101\\000\\010\end{bmatrix}=\frac{1}{7}$$

As predicted, $\rho\begin{bmatrix}0110\\1000\\1000\\000x\end{bmatrix}=0$ and the lower bound is refined to be -1.143, which matches the periodic upper bound. Thus, we prove that structure given in Figure 4b is the true ground state.

Although we have only demonstrated this algorithm using small 2D binary systems, we have successfully applied cluster tree optimization to automatically solve systems with basic block sizes up to 4 by 5. We have also successfully applied it to a 3D binary system with a block size up to 2 by 3 by 3. Finally, we have successfully generalized this algorithm to multicomponent cases, although demonstrating the details of these solutions is exceedingly tedious. In terms of computational complexity, the bottleneck of this algorithm is the initial enumerations of elements in the minimal block, where the minimal block is the smallest block to capture all interactions. The complexity order is thus $O(k^{x \cdot y \cdot z})$ where k is the numbers of components, and x, y, z is the minimal block size in the x, y, and z directions necessary to capture all interactions. As discussed earlier, while the complexity is exponential in the length scale of the interactions, the exponent is much smaller than that required for the traditional polytope method, making our algorithm much more tractable for solving realistic systems. While our current computational limit is $k^{x \cdot y \cdot z} < 2^{20}$, but this limit is not fundamental, and we intend to address methods to void the necessity to enumerate basic blocks in future work.

## Conclusion

We have presented a method for obtaining the ground state of a generalized Ising model by the novel cluster tree optimization algorithm. We have proven the correctness of this approach for finding periodic ground states, and shown that even when a periodic ground state solution cannot be found, this algorithm provides a sequence of states with energy converging to ground state energy.

Our approach voids the necessity of exponentially-difficult enumeration to counter frustration. Thus it enables us to probe the space of ground states by directly enumerating the vertices in the true polytope, automatically eliminating unconstructible vertices.

# Supplementary Information:

## *Proof of the undecidability of the ground state problem*

Suppose there exists an algorithm that, given arbitrary ECI, is guaranteed to produce the ground state configuration and ground state energy of a generalized Ising model/cluster expansion. Now consider arbitrary set of corner Wang tiles. We define the ECI such that all block energies corresponding to an element inside the set of Wang tiles to be -1 and all block energies corresponding to an element outside the set to be 0. Now, input this set of ECI into the presupposed algorithm. We could then get the ground state energy and ground state configuration. If the ground state energy is larger than -1, we can conclude that the set of tiles could not tile the plane. Otherwise, the ground state energy is -1, and we have the ground state spin configuration. Using arguments analogous to [47], we could show that there exists a tiling composed of only elements in the tile set and thus the tile set could tile the plane. Thus, the algorithm to calculate ground state corresponding to the given ECI can be modified to decide whether a given set of Wang tile can tile the plane, violating the undecidability of the Wang tile problem. Thus, the ground state problem must be undecidable. ∎

## *Proof of the equivalence between a solution obtained on an orthorhombic with no symmetry and that obtained on a general lattice*

Firstly, note that we can always construct a bijection from all configurations on the motif with n sites to integers ranging from 1 to $2^n$. Indeed, we could easily extend the bijection from a binary system to m-nary (binary, ternary, quaternary, etc.) system.

All that remains is to show that all interactions on an arbitrary 2D or 3D lattice could be interpreted as interactions on an orthorhombic lattice. Within some configuration, every energy term is written as: $\sigma_i \sigma_j \cdots$, with **i, j** being **2D (3D) vectors** denoting the position of the spin. Every positional vector of the spin is an integral sum of primitive vectors of the lattice and the corresponding integer vectors **n, m**. Thus this energy term could be written as $\sigma_n \sigma_m \cdots$. We could now rewrite the spin position **n, m** as on an orthorhombic lattice. ∎

## *Examples of transforming an arbitrary lattice system to an orthorhombic lattice*

**BCC System to Simple Orthorhombic:**

We will illustrate how to view a bcc lattice with nearest body interaction and next nearest body interaction in terms of interactions within a cube.

All the vectors denoting nearest body interaction is:

$$\left(\frac{1}{2},\frac{1}{2},\frac{1}{2}\right), \left(-\frac{1}{2},\frac{1}{2},\frac{1}{2}\right), \left(\frac{1}{2},-\frac{1}{2},\frac{1}{2}\right), \left(\frac{1}{2},\frac{1}{2},-\frac{1}{2}\right), \left(-\frac{1}{2},-\frac{1}{2},\frac{1}{2}\right), \left(-\frac{1}{2},\frac{1}{2},-\frac{1}{2}\right),$$
$$\left(\frac{1}{2},-\frac{1}{2},-\frac{1}{2}\right), \left(-\frac{1}{2},-\frac{1}{2},-\frac{1}{2}\right)$$

By defining the second, third and forth terms to be the primitive vector, we could represent the above vectors as:

$$[1,1,1],[1,0,0],[0,1,0],[0,0,1],[0,0,-1],[0,-1,0],[-1,0,0],[-1,-1,-1]$$

For the next nearest neighbor, the vectors are:

$$(1,0,0),(-1,0,0),(0,1,0),(0,-1,0),(0,0,1),(0,0,-1)$$

They could be represented using the primitive vector as:

$$[0,1,1],[0,-1,-1],[1,0,1],[-1,0,-1],[1,1,0],[-1,-1,0]$$

So the Hamiltonian could be exactly reproduced as:

$$H = \sum_{i\in\{\text{spin sites}\}} \sum_{N\in\{\text{nearest neighbor in bcc}\}} J_N \sigma_i \sigma_{i+N} + \sum_{i\in\{\text{spin sites}\}} \sum_{NN\in\{\text{next nearest neighbor in bcc}\}} J_{NN} \sigma_i \sigma_{i+NN}$$
$$= \sum_{i\in\mathbb{Z}^3} \sum_{N\in\Omega_N} J_N \sigma_i \sigma_{i+N} + \sum_{i\in\mathbb{Z}^3} \sum_{NN\in\Omega_{NN}} J_{NN} \sigma_i \sigma_{i+NN}$$

With

$$\{\text{spin sites}\} = \left\{\mathbf{v}: \mathbf{v} = i\cdot\left(-\frac{1}{2},\frac{1}{2},\frac{1}{2}\right) + j\cdot\left(\frac{1}{2},-\frac{1}{2},\frac{1}{2}\right) + k\cdot\left(\frac{1}{2},\frac{1}{2},-\frac{1}{2}\right), i,j,k \in \mathbb{Z}^3\right\}$$

$$\{\text{nearest neighbor in bcc}\} = \left\{\begin{array}{l}\left(\frac{1}{2},\frac{1}{2},\frac{1}{2}\right),\left(-\frac{1}{2},\frac{1}{2},\frac{1}{2}\right),\left(\frac{1}{2},-\frac{1}{2},\frac{1}{2}\right),\left(\frac{1}{2},\frac{1}{2},-\frac{1}{2}\right),\\ \left(-\frac{1}{2},-\frac{1}{2},\frac{1}{2}\right),\left(-\frac{1}{2},\frac{1}{2},-\frac{1}{2}\right),\left(\frac{1}{2},-\frac{1}{2},-\frac{1}{2}\right),\left(-\frac{1}{2},-\frac{1}{2},-\frac{1}{2}\right)\end{array}\right\}$$

$$\{\text{next nearest neighbor in bcc}\} = \{(1,0,0),(-1,0,0),(0,1,0),(0,-1,0),(0,0,1),(0,0,-1)\}$$

$$\Omega_N = \{[1,1,1],[1,0,0],[0,1,0],[0,0,1],[0,0,-1],[0,-1,0],[-1,0,0],[-1,-1,-1]\}$$
$$\Omega_{NN} = \{[0,1,1],[0,-1,-1],[1,0,1],[-1,0,-1],[1,1,0],[-1,-1,0]\}$$

As a result, we see that the minimization over the bcc lattice is exactly equivalent to the minimization over the cubic lattice without symmetry. ∎

**FCC Lattice to Simple Orthorhombic:**

We examine again the conversion from fcc lattice to cubic lattice. The procedure is exactly the same as above:

List nearest neighbor vectors:

$$\left(\frac{1}{2},\frac{1}{2},0\right), \left(-\frac{1}{2},\frac{1}{2},0\right), \left(\frac{1}{2},-\frac{1}{2},0\right), \left(-\frac{1}{2},-\frac{1}{2},0\right), \left(\frac{1}{2},0,\frac{1}{2}\right), \left(-\frac{1}{2},0,\frac{1}{2}\right)$$

$$\left(\frac{1}{2},0,-\frac{1}{2}\right), \left(-\frac{1}{2},0,-\frac{1}{2}\right), \left(0,\frac{1}{2},\frac{1}{2}\right), \left(0,-\frac{1}{2},\frac{1}{2}\right), \left(0,\frac{1}{2},-\frac{1}{2}\right), \left(0,-\frac{1}{2},-\frac{1}{2}\right)$$

Define the primitive vector:

$$\left(\frac{1}{2},\frac{1}{2},0\right) \to [0,0,1]$$

$$\left(\frac{1}{2},0,\frac{1}{2}\right) \to [0,1,0]$$

$$\left(0,\frac{1}{2},\frac{1}{2}\right) \to [1,0,0]$$

The representation in cube of the nearest neighbors in fcc is:

$$[0,0,1],[1,-1,0],[-1,1,0],[0,0,-1],[0,1,0],[1,0,-1],$$
$$[-1,0,1],[0,-1,0],[1,0,0],[0,1,-1],[0,-1,1],[-1,0,0]$$

The next nearest neighbor list in fcc is:

$$(1,0,0),(-1,0,0),(0,1,0),(0,-1,0),(0,0,1),(0,0,-1)$$

The corresponding representation in cube is:

$$[-1,1,1],[1,-1,-1],[1,-1,1],[-1,1,-1],[1,1,-1],[-1,-1,1]$$

We notice for both fcc and bcc, the nearest body and next nearest body all lie within the range of a cube of size 2 by 2 by 2. ∎

### Proof of the perfect sum relationship between the appearing frequencies of blocks and subblocks:

We prove this relationship by mathematical induction. In the base case N=n, eq. 7 is clearly correct. Now supposing that this relationship holds for N, we now show that eq. 7 also holds for N+1:

Note that when we have constructed equivalent forms of equations 1, 2, 3, and 4 for blocks of size N+1 by N+1, we have also spawned the blocks appearing in the right hand side of equations 2, 3 and 4 for all nonzero blocks of size N+1 by N+1. Since the blocks appearing on the right hand side of equations 2, 3, and 4 are necessarily non-zero, we have also spawned them into the corresponding N+1 by N+1 blocks. The spawning procedure insures the **perfect sum** relationship between blocks of size N and size N+1. Originally, we have inductively assumed a **perfect sum** relationship between blocks of size n and size N. Thus, the **perfect sum** relationship between blocks of size n and size N+1 follows by induction. ∎

### Proof of the constructability halting criterion:

Note that the perfect sum relationship derived earlier holds for blocks of size N and N-1:

$$\rho\left[\mathbf{B}_{N-1,N-1}\right] = \sum_{\mathbf{x}} \rho\begin{bmatrix} \mathbf{x}_\alpha & x_b \\ \mathbf{B}_{N-1,N-1} & \mathbf{x}_\beta \end{bmatrix}$$

The condition that every non-zero blocks of size N by N admits the same periodicity requires that $\mathbf{B}_{N-1,N-1}$ uniquely determines $\begin{bmatrix} \mathbf{x}_\alpha & x_b \\ \mathbf{B}_{N-1,N-1} & \mathbf{x}_\beta \end{bmatrix}$. Thus, there exists one unique $(\mathbf{x}_\alpha, x_b, \mathbf{x}_\beta)$ such that:

$$\rho\left[\mathbf{B}_{N-1,N-1}\right] = \rho\begin{bmatrix} \mathbf{x}_\alpha & x_b \\ \mathbf{B}_{N-1,N-1} & \mathbf{x}_\beta \end{bmatrix} \quad \textbf{(Eq. 11)}$$

And similarly for all three other directions:

$$\rho\left[\mathbf{B}_{N-1,N-1}\right] = \rho\begin{bmatrix} \mathbf{y}_\alpha & \mathbf{B}_{N-1,N-1} \\ y_b & \mathbf{y}_\beta \end{bmatrix} \quad \textbf{(Eq. 12)}$$

$$\rho\left[\mathbf{B}_{N-1,N-1}\right] = \rho\begin{bmatrix} \mathbf{B}_{N-1,N-1} & \mathbf{x}_\beta \\ \mathbf{y}_\beta & z_b \end{bmatrix} \quad \textbf{(Eq. 13)}$$

$$\rho\begin{bmatrix} \mathbf{B}_{N-1,N-1} \end{bmatrix} = \rho\begin{bmatrix} w_b & \mathbf{x}_\alpha \\ \mathbf{y}_\alpha & \mathbf{B}_{N-1,N-1} \end{bmatrix}$$ **(Eq. 14)**

Given $\mathbf{B}_{N-1,N-1}$ and a fixed periodicity, a global configuration is uniquely determined. Taking every block of size $(N-1, N-1)$ and $(N, N)$ in this global configuration and apply equation (Eq. 11), (Eq. 12), (Eq. 13) and (Eq. 14), we realize all of them have the same appearing frequency. If all other block appearing frequency of size $(N-1, N-1)$ is 0, then the appearing frequency vector corresponds to the global configuration. If not, by setting all the previously related N by N blocks and N-1 by N-1 blocks to be 0 and repeating the procedure above, we arrive at another global configuration with appearing frequency of every $(N-1, N-1)$ and $(N, N)$ blocks being the same. It could be then realized that this appearing frequency vector is a sum of constructible structures and thus it is constructible. ∎

## Rigorous definition of spawning

In our algorithm, we have introduced the term **spawn**, which is not in this current form exact enough for one to reproduce all of our procedures. In general, **spawning** is simply a way to introduce equalities and inequalities to the linear programming system.

Assume the interaction range is m by n, defined previously in the manuscript. As before, the variables defined in the compatibility equation are:

$$\rho\begin{bmatrix} \sigma_{1,1} & \cdots & \sigma_{1,m} \\ \vdots & \ddots & \vdots \\ \sigma_{n,1} & \cdots & \sigma_{n,m} \end{bmatrix}$$

Thus, spawning can be rigorously defined as:

(1) Spawning on a variable of the form $\rho\begin{bmatrix} \sigma_{1,1} & \cdots & \cdots & \sigma_{1,M} \\ \vdots & \ddots & \ddots & \vdots \\ \vdots & \ddots & \ddots & \vdots \\ \sigma_{N,1} & \cdots & \cdots & \sigma_{N,M} \end{bmatrix}$ (with $N - n = M - m$) means to introduce the following equations into our linear programming system:

$$\rho\begin{bmatrix} \sigma_{1,1} & \cdots & \sigma_{1,M} \\ \vdots & \ddots & \vdots \\ \vdots & \ddots & \vdots \\ \sigma_{N,1} & \cdots & \sigma_{N,M} \end{bmatrix} = \sum_s \rho\begin{bmatrix} \sigma_{1,1} & \cdots & \sigma_{1,M} \\ \vdots & \ddots & \vdots & s_1 \\ \vdots & \ddots & \vdots & \vdots \\ \sigma_{N,1} & \cdots & \sigma_{N,M} & s_n \end{bmatrix} = \sum_s \rho\begin{bmatrix} & \sigma_{1,1} & \cdots & \sigma_{1,M} \\ s_1 & \vdots & \ddots & \vdots \\ \vdots & \vdots & \ddots & \vdots \\ s_n & \sigma_{N,1} & \cdots & \sigma_{N,M} \end{bmatrix}$$

$$= \sum_s \rho\begin{bmatrix} \sigma_{1,1} & \cdots & \sigma_{1,M} & s_1 \\ \vdots & \ddots & \vdots & \vdots \\ \vdots & \ddots & \vdots & s_n \\ \sigma_{N,1} & \cdots & \sigma_{N,M} & \end{bmatrix} = \sum_s \rho\begin{bmatrix} s_1 & \sigma_{1,1} & \cdots & \sigma_{1,M} \\ \vdots & \vdots & \ddots & \vdots \\ s_n & \vdots & \ddots & \vdots \\ & \sigma_{N,1} & \cdots & \sigma_{N,M} \end{bmatrix}$$

$$\forall s \; \rho\begin{bmatrix} \sigma_{1,1} & \cdots & \cdots & \sigma_{1,M} \\ \vdots & \ddots & \ddots & \vdots & s_1 \\ \vdots & \ddots & \ddots & \vdots & \vdots \\ \sigma_{N,1} & \cdots & \cdots & \sigma_{N,M} & s_n \end{bmatrix} \leq \rho\begin{bmatrix} \sigma_{N-n+1,M+2-m} & \cdots & \sigma_{N-n+1,M} & s_1 \\ \vdots & \ddots & \vdots & \vdots \\ \sigma_{N,M+2-m} & \cdots & \sigma_{N,M} & s_n \end{bmatrix}$$

$$\forall s \; \rho\begin{bmatrix} & \sigma_{1,1} & \cdots & \cdots & \sigma_{1,M} \\ s_1 & \vdots & \ddots & \ddots & \vdots \\ \vdots & \vdots & \ddots & \ddots & \vdots \\ s_n & \sigma_{N,1} & \cdots & \cdots & \sigma_{N,M} \end{bmatrix} \leq \rho\begin{bmatrix} s_1 & \sigma_{N-n+1,1} & \cdots & \sigma_{N-n+1,m-1} \\ \vdots & \ddots & \vdots & \vdots \\ s_n & \sigma_{N,1} & \cdots & \sigma_{N,m-1} \end{bmatrix}$$

(2) Spawning on a variable of the form $\rho\begin{bmatrix} \sigma_{1,1} & \cdots & \cdots & \sigma_{1,M} \\ \vdots & \ddots & \ddots & \vdots & \mu_1 \\ \vdots & \ddots & \ddots & \vdots & \vdots \\ \sigma_{N,1} & \cdots & \cdots & \sigma_{N,M} & \mu_n \end{bmatrix}$ (with $N - n = M - m$) means to introduce these following equations into our linear programming system:

$$\rho\begin{bmatrix} \sigma_{1,1} & \cdots & \cdots & \sigma_{1,M} \\ \vdots & \ddots & \ddots & \vdots & \mu_1 \\ \vdots & \ddots & \ddots & \vdots & \vdots \\ \sigma_{N,1} & \cdots & \cdots & \sigma_{N,M} & \mu_n \end{bmatrix} = \sum_s \rho\begin{bmatrix} s_1 & \cdots & s_m & & \\ \sigma_{1,1} & \cdots & \cdots & \sigma_{1,M} \\ \vdots & \ddots & \ddots & \vdots & \mu_1 \\ \vdots & \ddots & \ddots & \vdots & \vdots \\ \sigma_{N,1} & \cdots & \cdots & \sigma_{N,M} & \mu_n \end{bmatrix}$$

$$\forall s \quad \rho \begin{bmatrix} s_1 & \cdots & s_m & & \\ \sigma_{1,1} & \cdots & \cdots & \sigma_{1,M} & \\ \vdots & \ddots & \ddots & \vdots & \mu_1 \\ \vdots & \ddots & \ddots & \vdots & \vdots \\ \sigma_{N,1} & \cdots & \cdots & \sigma_{N,M} & \mu_n \end{bmatrix} \leq \rho \begin{bmatrix} s_1 & \cdots & s_m \\ \sigma_{1,1} & \cdots & \sigma_{1,m} \\ \vdots & \ddots & \vdots \\ \sigma_{n-1,1} & \ddots & \sigma_{n-1,m} \end{bmatrix}$$

(3) Spawning on a variable of the form $\rho \begin{bmatrix} \psi_1 & \cdots & \psi_\alpha & & \\ \sigma_{1,1} & \cdots & \cdots & \sigma_{1,M} & \\ \vdots & \ddots & \ddots & \vdots & \mu_1 \\ \vdots & \ddots & \ddots & \vdots & \vdots \\ \sigma_{N,1} & \cdots & \cdots & \sigma_{N,M} & \mu_\beta \end{bmatrix}$ (with

$N - n = M - m$) means to introduce these following equations into our linear programming system:

**If** $\beta - n > \alpha - m$:

$$\rho \begin{bmatrix} \psi_1 & \cdots & \psi_\alpha & & \\ \sigma_{1,1} & \cdots & \cdots & \sigma_{1,M} & \\ \vdots & \ddots & \ddots & \vdots & \mu_1 \\ \vdots & \ddots & \ddots & \vdots & \vdots \\ \sigma_{N,1} & \cdots & \cdots & \sigma_{N,M} & \mu_\beta \end{bmatrix} = \sum_k \rho \begin{bmatrix} \psi_1 & \cdots & \psi_\alpha & k & & \\ \sigma_{1,1} & \cdots & \cdots & \cdots & \sigma_{1,M} & \\ \vdots & \ddots & \ddots & \cdots & \vdots & \mu_1 \\ \vdots & \ddots & \ddots & \cdots & \vdots & \vdots \\ \sigma_{N,1} & \cdots & \cdots & \cdots & \sigma_{N,M} & \mu_\beta \end{bmatrix}$$

$$\forall k \quad \rho \begin{bmatrix} \psi_1 & \cdots & \psi_\alpha & k & & \\ \sigma_{1,1} & \cdots & \cdots & \cdots & \sigma_{1,M} & \\ \vdots & \ddots & \ddots & \cdots & \vdots & \mu_1 \\ \vdots & \ddots & \ddots & \cdots & \vdots & \vdots \\ \sigma_{N,1} & \cdots & \cdots & \cdots & \sigma_{N,M} & \mu_\beta \end{bmatrix} \leq \rho \begin{bmatrix} \psi_1 & \cdots & \psi_\alpha & k \\ \sigma_{1,1} & \cdots & \cdots & \sigma_{1,\alpha+1} \\ \vdots & \ddots & \ddots & \cdots \\ \vdots & \ddots & \ddots & \cdots \\ \sigma_{n+\alpha-m,1} & \cdots & \cdots & \sigma_{n+\alpha-m,\alpha+1} \end{bmatrix}$$

**If** $\beta - n = \alpha - m$

$$\rho\begin{bmatrix} \psi_1 & \cdots & \psi_\alpha & & & & \\ \sigma_{1,1} & \cdots & \cdots & \sigma_{1,M} & & & \\ \vdots & \ddots & \ddots & \vdots & & & \mu_1 \\ \vdots & \ddots & \ddots & \vdots & & & \vdots \\ \sigma_{N,1} & \cdots & \cdots & \sigma_{N,M} & & & \mu_\beta \end{bmatrix} = \sum_k \rho\begin{bmatrix} \psi_1 & \cdots & \psi_\alpha & & & & \\ \sigma_{1,1} & \cdots & \cdots & \sigma_{1,M} & & & \\ \vdots & \vdots & \vdots & \vdots & & & k \\ & & & & & & \mu_1 \\ \vdots & \ddots & \ddots & \vdots & & & \vdots \\ \sigma_{N,1} & \cdots & \cdots & \sigma_{N,M} & & & \mu_\beta \end{bmatrix}$$

$$\forall k \quad \rho\begin{bmatrix} & \psi_1 & \cdots & \psi_\alpha & & \\ \sigma_{1,1} & \cdots & \cdots & \sigma_{1,M} & & \\ \vdots & \vdots & \vdots & \vdots & & k \\ & & & & & \mu_1 \\ \vdots & \ddots & \ddots & \vdots & & \vdots \\ \sigma_{N,1} & \cdots & \cdots & \sigma_{N,M} & & \mu_\beta \end{bmatrix} \leq \rho\begin{bmatrix} \sigma_{N-\beta,M+1-\beta+n-m} & \cdots & \sigma_{N-\beta,M} & k \\ \vdots & \ddots & \vdots & \mu_1 \\ \vdots & \ddots & \vdots & \vdots \\ \sigma_{N,M+1-\beta+n-m} & \cdots & \sigma_{N,M} & \mu_\beta \end{bmatrix}$$

(4) Similarly we define spawning on variables of the form

$$\rho\begin{bmatrix} & \sigma_{1,1} & \cdots & \cdots & \sigma_{1,M} \\ \mu_1 & \vdots & \ddots & \ddots & \vdots \\ \vdots & \vdots & \ddots & \ddots & \vdots \\ \mu_n & \sigma_{N,1} & \cdots & \cdots & \sigma_{N,M} \end{bmatrix}, \rho\begin{bmatrix} \mu_1 & \sigma_{1,1} & \cdots & \cdots & \sigma_{1,M} \\ \vdots & \vdots & \ddots & \ddots & \vdots \\ \mu_n & \vdots & \ddots & \ddots & \vdots \\ & \sigma_{N,1} & \cdots & \cdots & \sigma_{N,M} \end{bmatrix},$$

$$\rho\begin{bmatrix} \sigma_{1,1} & \cdots & \cdots & \sigma_{1,M} & \mu_1 \\ \vdots & \ddots & \ddots & \vdots & \vdots \\ \vdots & \ddots & \ddots & \vdots & \mu_n \\ \sigma_{N,1} & \cdots & \cdots & \sigma_{N,M} & \end{bmatrix}$$ as first, applying a mirror symmetry on $\rho$ to obtain a

block in the form of (2), generating equations according to (2) and finally reversing the mirror symmetry operation on every term generated in these equations and taking those equations as new introduced constraints into the linear programming system.

(5) Similarly we define spawning on variables of the form,

$$\rho\begin{bmatrix} & & \psi_1 & \cdots & \psi_\alpha & & \\ & \sigma_{1,1} & \cdots & \cdots & \sigma_{1,M} & \\ \mu_1 & \vdots & \ddots & \ddots & \vdots & \\ \vdots & \vdots & \ddots & \ddots & \vdots & \\ \mu_\beta & \sigma_{N,1} & \cdots & \cdots & \sigma_{N,M} & \end{bmatrix}, \rho\begin{bmatrix} \mu_1 & \sigma_{1,1} & \cdots & \cdots & \sigma_{1,M} \\ \vdots & \vdots & \ddots & \ddots & \vdots \\ \mu_\beta & \vdots & \ddots & \ddots & \vdots \\ & \sigma_{N,1} & \cdots & \cdots & \sigma_{N,M} \\ & & \psi_1 & \cdots & \psi_\alpha \end{bmatrix},$$

$$\rho \begin{bmatrix} \sigma_{1,1} & \cdots & \cdots & \sigma_{1,M} & \mu_1 \\ \vdots & \ddots & \ddots & \vdots & \vdots \\ \vdots & \ddots & \ddots & \vdots & \mu_\beta \\ \sigma_{N,1} & \cdots & \cdots & \sigma_{N,M} & \\ \psi_1 & \cdots & \psi_\alpha & & \end{bmatrix}$$ as first applying a mirror symmetry on $\rho$ to obtain a block in the form of (3), generating equations according to (3) and finally reversing the mirror symmetry operation on every term generated in those equations and taking those equations as new introduced constraints into the linear programming system.